\documentclass[aps,twocolumn,PRA,reprint,superscriptaddress]{revtex4}
\usepackage{amsmath}
\usepackage[percent]{overpic}
\usepackage{times}
\usepackage{subfigure}
\usepackage{latexsym}
\usepackage{graphicx}
\usepackage{bm}
\usepackage{amssymb}
\usepackage{hyperref}
\usepackage{multirow}
\usepackage{lipsum}

\begin{document}


\title{Spin-orbit coupled spin-1 Bose-Einstein condensate flow past an obstacle in the presence of a Zeeman field}

\author{Qing-Li Zhu}
\affiliation{National Laboratory of Solid State Microstructures and Department of Physics, Nanjing University, Nanjing 210093, China}
\affiliation{Department of information engineering, Nanjing Normal University Taizhou college, Taizhou 225300, China}
\author{Lihua Pan}
\affiliation{School of Physics Science and Technology, Yangzhou University, Yangzhou 225002, China}
\author{Jin An}
\email{anjin@nju.edu.cn}
\affiliation{National Laboratory of Solid State Microstructures and Department of Physics, Nanjing University, Nanjing 210093, China}
\affiliation{Collaborative Innovation Center of Advanced Microstructures, Nanjing University, Nanjing 210093, China}
\date{\today}

\begin{abstract}
We study the dynamics of a Rashba spin-orbit coupled spin-1 ferromagnetic Bose-Einstein condensate under a linear Zeeman magnetic field(ZF) disturbed by a moving obstacle. The Bogoliubov excitation spectrums and corresponding critical excitations in different situations are analyzed. The structure of the coreless vortex or antivortex generated by the moving obstacle has been investigated. When the ZF is applied along $\bm{x}$ direction, the vortex cores for the three components of a(an) vortex(antivortex) could be arranged into a vertical line, and their order would be reversed as the spin-orbit coupling increases. When the ZF is parallel to $\bm{z}$ direction, a skyrmion-like vortex ground state could be induced even by a static obstacle. This topological structure is also found to be dynamically stable if the obstacle is moving at a relatively small velocity.
\end{abstract}

\maketitle

\section{INTRODUCTION}

Since the realization of synthetic non-Abelian gauge fields in Bose-Einstein condensate (BEC) by coupling the internal spin states and orbital momentum of atoms\cite{lin}, the spinor BECs have attracted much attention by not merely providing an ideal platform to simulate the spin-orbit (SO) coupling effect in condensed matter\cite{spinhall,qxl,hasan}, but also exhibiting plenty of exotic phenomena in cold-atom systems. Apart from various novel ground states such as half-quantum vortex \cite{wucj,sinha,Ramachandhran,xxq1}, stripe phase \cite{zhai1,ho2,sinha,martone,putra,zyp,zqyu,ksun,wjg}and vortex related structures\cite{xzf,kato1}, the introduction of SO coupling has led to many other possibilities of topological textures including skyrmions \cite{kawakami,lcf2012,xzf,gjchen}, meron \cite{wilson,zxf2012,dongb} and monopoles \cite{monopole,liji2017}. Recently, relevant work has been extended to SO induced supersolid phase\cite{lijr,luoxw,solid2020}, spin-tensor-momentum coupling\cite{tensor} , spin-nematic-orbit coupling \cite{nematic} and another fundamental type, namely, the coupling between spin and orbital angular momentum of atoms \cite{demarco,sunk}, where phenomena like the splitting of vortex cores \cite{chr} and the first-order phase transitions\cite{zdf} have been observed experimentally.

Dynamical properties also play an important role in characterizing spinor BECs. In the past decade, various topological collective excitations including exotic vortex or vortex pair \cite{love2014,seosw,borgh, kangs,love2012,fetter2014, williamson,kato2}, soliton\cite{achil,dark,bright,emerson,gautam,jiacl,sunjie}, knot\cite{dshall,ollika, ykliu}, skyrmion\cite{savage,choijy,susw,lcf1,ollika2,ivana} have been proposed and their dynamical stability have either been theoretically discussed or experimentally verified in the framework of spinor BECs. Recently, a variety of studies were performed on the dynamics of a scalar BEC flow past an obstacle, especially after the experimental observation of the induced vortex-antivortex pairs\cite{neely,freilich}, which have been shown to exhibit extraordinary behaviors\cite{sasaki,kwona,aioi2011,fujimoto,pinsker,kadokura,kunimi,kwonb,katsi}. Nevertheless, much less attention has been paid on the corresponding case of a spinor BEC flow past an obstacle\cite{rodr,khamehchi,yli,kato3}, which is expected to be capable of revealing more exotic quantum states due to the interplay among SO coupling, spin exchange and other competing interactions.

In this paper, we are focused on the dynamical problem of a spin-1 ferromagnetic BEC flow past an obstacle. By emphasizing the competition between the Rashba SO coupling and the linear ZF, we find the structure of the vortex or antivortex generated by the obstacle shows novel feature. In the case of ZF being along $\bm{x}$ direction, the vortex cores for the three components line up vertically, and their order would be reversed as the SO coupling increases, while in the case of ZF being along $\bm{z}$ direction, even a static obstacle could induce a skyrmion-like vortex ground state, whose topological structure is also found to be dynamically stable if the obstacle is moving at a relatively small velocity.

This paper is organized as follows. In the next section, we introduce our model system and discuss its mean-field ground states under different conditions. In sec.III, we analyze in detail its Bogoliubov excitation spectrum, discuss the critical velocity of the moving obstacle, as well as the corresponding critical excitations(CEs). In Sec.IV, by numerically solving the Gross-Pitaevskii(GP) equation, we investigate the vortices(antivortices) generated by the obstacle, and analyze their vortex structures and topologies in different situations. In Sec.V, we summarize our results.

\section{MODEL}

Equivalently, we consider a static homogeneous quasi-two-dimensional spin-1 BEC with SO coupling, in which a moving obstacle is passing by. In consideration of the linear Zeeman effect, the Hamiltonian of such a system is given by $H = H_{0}+H_{\text{int}}+H_{\text{obstacle}}$,
\begin{equation}
\begin{split}
&H_{0}=\int d\bm{r}\Psi^{\dagger}[\frac{-\hbar^2\bm{\nabla}^2}{2m}+\nu_{soc}+g\mu_{B}\bm{B}\cdot\bm{\hat{F}}]\Psi\\
&H_{\text{int}}=\int d\bm{r}(\frac{1}{2}c_{0}n^{2}+\frac{1}{2}c_{2}\langle\hat{\bm{F}}\rangle^{2})\\
&H_{\text{obstacle}}=V(\bm{r}-\bm{v}t),
\end{split}
\end{equation}
where $\Psi=[\psi_{1}(\bm{r}),\psi_{0}(\bm{r}),\psi_{-1}(\bm{r})]^{T}$ denotes the spinor order parameter and is normalized to satisfy
$\int d\bm{r}\Psi^{\dagger}\Psi=N$. Here, $n=|\psi_{1}|^{2}+|\psi_{0}|^{2}+|\psi_{-1}|^{2}$ is atomic density, while $\langle\hat{\bm{F}}\rangle=\Psi^{\dagger}\bm{\hat{F}}\Psi$ is spin density with $\bm{\hat{F}}=(F_{x},F_{y},F_{z})$ being spin-1 pauli matrices. For the SO interaction, we consider the Rashba coupling with $\nu_{\text{soc}}=\frac{\hbar k_{0}}{m}(-i\hbar)(F_{x}\partial_{x}+F_{y}\partial_{y})$, in which $k_{0}$ denotes its strength. The linear ZF is assumed to be in the $x-z$ plane, and its strength is represented by $g\mu_{B}B\equiv p_{0}$. $H_{\text{int}}$ denotes the standard contact and spin-exchange interactions, where the latter one favors the ferromagnetic(FM) ground state for $c_{2}<0$ and polar ground state for $c_{2}>0$. In this paper, we focus on the FM case. $H_{\text{obstacle}}$ is the potential of the moving obstacle with constant velocity $\bm{v}$, which takes the circular form,
\begin{equation}
V(\bm{r})=\left\{
\begin{array}{rcl}
V_{0} &   &{|\bm{r}|\leq R}\\
0     &   &{|\bm{r}|\ge  R},
\end{array} \right.
\end{equation}
where the potential height $V_{0} $ is taken to be much larger than the chemical potential $\mu$.

For a homogeneous gas, it is instructive to start our investigation with the noninteracting case in the absence of the obstacle. When the ZF is taken along $\bm{x}$ direction, the single-particle Hamiltonian in $\bm{k}$-space can be given by
\begin{equation}
\begin{split}
H_{0}(\bm{k})=
&\left(
\begin{array}{ccc}
\frac{\hbar^{2}k^{2}}{2m}     &\frac{\hbar^{2}k_{0}(k_{-} + k_{p})}{\sqrt{2}m}              & 0  \\
\frac{\hbar^{2}k_{0}(k_{+} + k_{p})}{\sqrt{2}m}  &\frac{\hbar^{2}k^{2}}{2m}   &\frac{\hbar^{2}k_{0}(k_{-} + k_{p})}{\sqrt{2}m} \\
0                                    & \frac{\hbar^{2}k_{0}(k_{+} + k_{p})}{\sqrt{2}m}      &\frac{\hbar^{2}k^{2}}{2m}
\end{array}
\right ).
\end{split}
\end{equation}
Here $k_{p}=mp_{0}/\hbar^{2} k_{0}$, $k_{\pm}=k_{x}\pm ik_{y}$. Diagonalization of $H_{0}(\bm{k})$ leads to three energy bands,
\begin{equation}
E^{0}_{\bm{k}}=\frac{\hbar^{2}k^{2}}{2m},E^{\pm}_{\bm{k}}=\frac{\hbar^{2}k^{2}}{2m}\pm\frac{\hbar^{2}k_{0}}{m}\sqrt{k_{y}^{2}+(k_{x}+k_{p})^{2}}.
\end{equation}
When $\bm{k}=(k_{0},0)$, $E^{-}_{\bm{k}}$ reaches the global minimum $E_{g}=-\frac{\hbar^{2}k_{0}^{2}}{2m}-p_{0}$. Accordingly, the single-particle ground state is
\begin{equation}
\Psi_{g}(\bm{r})= e^{ik_{0}x}\frac{1}{2}
\left (
\begin{array}{ccc}
&1\\
&-\sqrt{2}\\
&1
\end{array}
\right ),
\end{equation}
which is fully spin polarized along $-\bm{x}$ direction.

While for the case of the ZF being along $\bm{z}$ direction, as is discussed in Ref.\cite{wenl}, the three energy bands with different helicities are:
\begin{equation}
E^{0}_{\bm{k}}=\frac{\hbar^{2}k^{2}}{2m},E^{\pm}_{\bm{k}}=\frac{\hbar^{2}k^{2}}{2m}\pm\frac{\hbar^{2}k_{0}}{m}\sqrt{k^{2}+k_{p}^{2}} .
\end{equation}
When $k_{p}< k_{0}$, the single-particle ground states with $E_{g}= -\frac{\hbar^{2}}{2m}(k_{0}^{2}+k_{p}^{2})$ are partially spin polarized states
\begin{equation}
\Psi_{g}(\bm{r})= \frac{1}{2k_{0}^{2}}
\left (
\begin{array}{ccc}
&(k_{0}^{2}-k_{p}^{2})e^{-i\theta_{k}}\\
&-\sqrt{2(k_{0}^{4}-k_{p}^{4})} \\
&(k_{0}^{2}+k_{p}^{2})e^{i\theta_{k}}
\end{array}
\right )
e^{i\bm{k}\cdot\bm{r}},
\end{equation}
in which $|\bm{k}|=(k_{0}^{2}-k_{p}^{2})^{\frac{1}{2}}$, and $\tan\theta_{k}=k_{y}/k_{x}$. While if $k_{p}\geq k_{0}$, $\Psi_{g}(\bm{r})=(0,0,1)^{T}$ with $E_{g}=-p_{0}$, which is fully spin polarized along $-\bm{z}$ direction.

In an infinite system, the atomic density far away from the obstacle is a constant $n_{0}$. In the following, we measure the length, energy and time by $\xi =\hbar/\sqrt{mc_{0}n_{0}}$(the healing length), $c_{0}n_{0}$, and $\hbar/(c_{0}n_{0})$, respectively. The obstacle velocity $\bm{v}$ are hence normalized by sound velocity $v_{s}=\sqrt{c_{0}n_{0}/m}$ under this unit. In the frame of the moving potential at velocity $\bm{v}$, the dimensionless GP equation for our model becomes
\begin{equation}
\begin{split}
&i\frac{\partial\psi_{1}}{\partial t}=[-\frac{1}{2}\nabla^2+V(\bm{r})+ i\bm{v}\cdot\nabla+n+p'_{z}]\psi_{1}+\frac{p'_{x}}{\sqrt{2}}\psi_{0}\\
&-\frac{i\kappa}{\sqrt{2}}\partial_{-}\psi_{0}+\gamma(n_{1}+n_{0}-n_{-1})\psi_{1}+\gamma\psi_{0}^{2}\psi_{-1}^{*}\\
&i\frac{\partial\psi_{0}}{\partial t}=[-\frac{1}{2}\nabla^2+V(\bm{r})+i\bm{v}\cdot\nabla+n]\psi_{0}+\frac{p'_{x}}{\sqrt{2}}(\psi_{1}+\psi_{-1})\\ &-\frac{i\kappa}{\sqrt{2}}(\partial_{+}\psi_{1}+\partial_{-}\psi_{-1})+\gamma(n_{1}+n_{-1})\psi_{0}+2\gamma\psi_{1}\psi_{-1}\psi_{0}^{*}\\
&i\frac{\partial\psi_{-1}}{\partial t}=[-\frac{1}{2}\nabla^2+V(\bm{r})+i\bm{v}\cdot\nabla+n-p'_{z}]\psi_{-1}+\frac{p'_{x}}{\sqrt{2}}\psi_{0}\\
&-\frac{i\kappa}{\sqrt{2}}\partial_{+}\psi_{0}+\gamma(n_{-1}+n_{0}-n_{1})\psi_{-1}+\gamma\psi_{0}^{2}\psi_{1}^{*},
\end{split}
\end{equation}
where $\partial_{\pm}=\partial_{x}\pm i\partial_{y}$, $\gamma=c_{2}/c_{0}$ and $\kappa=k_{0}\xi$. $p^{'}_{z}=p^{'}\cos\theta_{B}$ and $p^{'}_{x}=p^{'}\sin\theta_{B}$ with $p^{'}=k_{0}k_{p}\xi^{2}=p_{0}/c_{0}n_{0}$ and $\theta_{B}$ being the angle of $\bm{B}$ taken with $\bm{z}$. In the following, we use $\gamma=-0.05$, and we are also only focused on the two extreme cases with $\theta_{B}=0$ or $\pi/2$, since the cases for the other directions of ZF can be expected straightforwardly from them.

\begin{figure*}[!htb]
\includegraphics[scale=0.67]{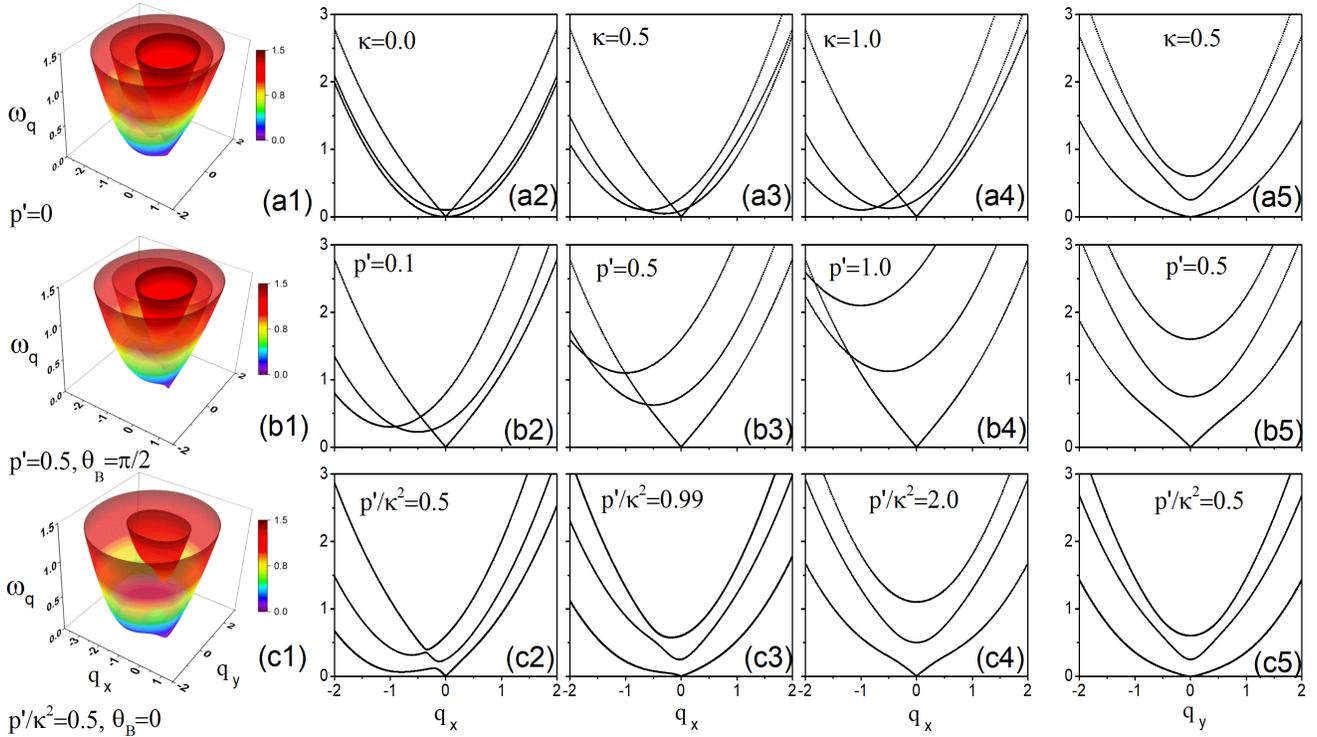}
\caption{(color online) Bogoliubov excitation spectrum of SO coupled spin-1 BEC in the presence of a linear ZF. From top to bottom the ZF is taken to be zero, along $\bm{x}$, $\bm{z}$ direction respectively. The left column indicates the 2D spectrums with SO coupling strength $\kappa=0.5$. The middle three columns and the right one denote respectively the spectrums scanning along $q_{x}$ and $q_{y}$ axis. Here, $\kappa=0.1$, $0.5$, $1.0$, $0.5$ for (a2) to (a5), and $p'=0.1$, $0.5$, $1.0$, $0.5$ for (b2) to (b5), while $p'/\kappa^{2}$ is taken as 0.5, 0.99, 2.0, 0.5 for (C2) to (C5) with $\kappa=0.5$. }
\end{figure*}

\section{results}
\subsection*{A: Analysis of Bogoliubov excitations}

We begin with an analysis on the Bogoliubov excitations in the moving frame without the obstacle. The wave function can be expanded as
\begin{equation}
\Psi(\bm{r},t)=e^{-i\mu t }
\left[\Psi_{g}(\bm{r}) + e^{i\tilde{\kappa}x}\delta\Psi(\bm{r},t)\right],
\end{equation}
where $\mu$ is the chemical potential, and $\tilde{\kappa}$ is the ground-state momentum. Thus we have
$i\hbar\partial_{t}\delta\Psi(\bm{r})= (\mathcal{H}+i\bm{v}\cdot\nabla)\delta\Psi(\bm{r})+ \mathcal{H}^{'}\delta\Psi^{*}(\bm{r})$. See Appendix for detail.

To obtain the excitation spectrum, $\delta\Psi(\bm{r},t)$ can be further expanded as
$\delta\Psi(\bm{r},t) =\chi_{1} e^{i(\bm{q}\cdot\bm{r}-\omega t)}+\chi_{2}^{*}e^{-i(\bm{q}\cdot\bm{r}-\omega^{*} t)}$,
with $\bm{q}$ and $\omega$ being the wave vector and frequency of excitation. For a homogeneous system, the excitation energy can be expressed as $\omega_{\bm{q}}-\bm{v}\cdot\bm{q}$, where $\omega_{\bm{q}}$ satisfies the Bogoliubov equation $\mathcal{H}(\bm{q})\chi= \omega_{\bm{q}}\chi$, with
\begin{equation}
\mathcal{H}(\bm{q})=
\left (
\begin{array}{ccc}
\mathcal{H}_{\bm{q}}       &  \mathcal{H}^{'}                            \\
-\mathcal{\mathcal{H}}^{'}                        &  -\mathcal{H}^{*}_{\bm{-q}}     \\
\end{array}
\right ),
\chi=\left (
\begin{array}{ccc}
\chi_{1}                \\
\chi_{2}
\end{array}
\right ).
\end{equation}
See Appendix for the detailed expressions of $\mathcal{H}_{\bm{q}}$ and $\mathcal{H}^{'}$.

The Bogoliubov equation is often used to study the stability characteristics of a stationary state. If there exists at least one complex eigenfrequency with nonzero $\mathrm{Im}(\omega_{\bm{q}})$, the state is dynamically unstable, while if an eigenfrequency with $\omega_{\bm{q}}<\bm{v}\cdot\bm{q}$ exists, the state is unstable thermodynamically in the moving frame. For a given moving direction $\phi_{\bm{v}}$, the Landau critical velocity $v^{c}(\phi_{\bm{v}})$ is the minimum value of velocity, i.e.,
$ v^{c}=\min\{\frac{\omega_{\bm{q}}}{\bm{q}\cdot\widehat{\bm{v}}}\}$ at which there exists an instability region with $\omega_{\bm{q}}<\bm{v}\cdot\bm{q}$ around the critical momentum $\bm{q}^{c}$( in polar coordinates, $\bm{q}^{c}\equiv(q^{c},\phi_{q^{c}})$).
\begin{figure}[!htb]
\includegraphics[scale=0.55]{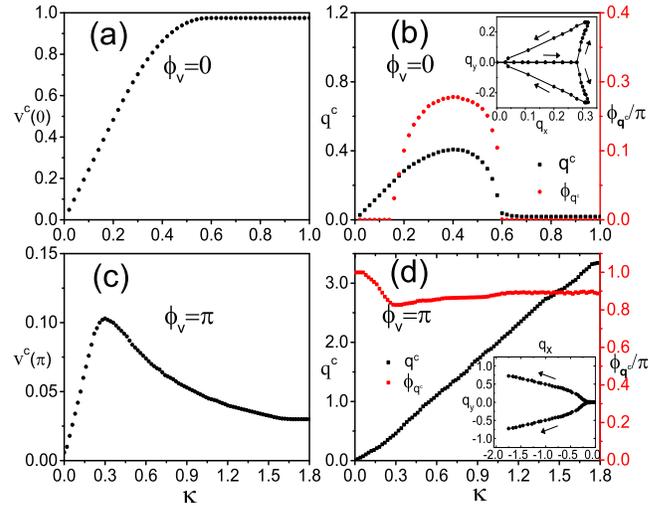}
\caption{(color online) Dependence of the critical velocity $v^{c}(\phi_{\bm{v}}=0)$, $v^{c}(\phi_{\bm{v}}=\pi)$ and the corresponding critical excitation momentum $\bm{q}^{c}$ on the SO coupling strength in the absence of ZF. The insets show the moving trajectories of $\bm{q}^{c}s$, where the arrows denote their moving directions as $\kappa$ increases. }
\end{figure}

In the case of $\theta_{B}=\frac{\pi}{2}$, if $\bm{q}^{c}=0$, after a lengthy derivation, an analytical result on $v^{c}(\phi_{\bm{v}})$ can be obtained,
\begin{equation}
v^{c}(\phi_{\bm{v}})= \min_{\phi_{\bm{q}}}\{\sqrt{(1+\gamma)\frac{\kappa^{2}{cos^{2}\phi_{\bm{q}}}+p'}{(\kappa^{2}+p')cos^{2}(\phi_{\bm{q}}-\phi_{\bm{v}})}}\}.
\end{equation}
Thus in the absence of ZF, $v^{c}(\phi_{\bm{v}})$ is independent of SO coupling and is given by
\begin{equation}
v^{c}(\phi_{\bm{v}})= \min_{\phi_{\bm{q}}}\{\sqrt{1+\gamma}|\frac{cos\phi_{\bm{q}}}{cos(\phi_{\bm{q}}-\phi_{\bm{v}})}|\}.
\end{equation}
Only as $\phi_{\bm{v}}=0$ or $\pi$, has $v^{c}$ a finite value $\sqrt{1+\gamma}$. Otherwise, $v^{c}=0$. Note that these analytical results are exact and valid only if the CE is located at $\bm{q}^{c}=0$.

\begin{figure*}[!htb]
\includegraphics[scale=0.62]{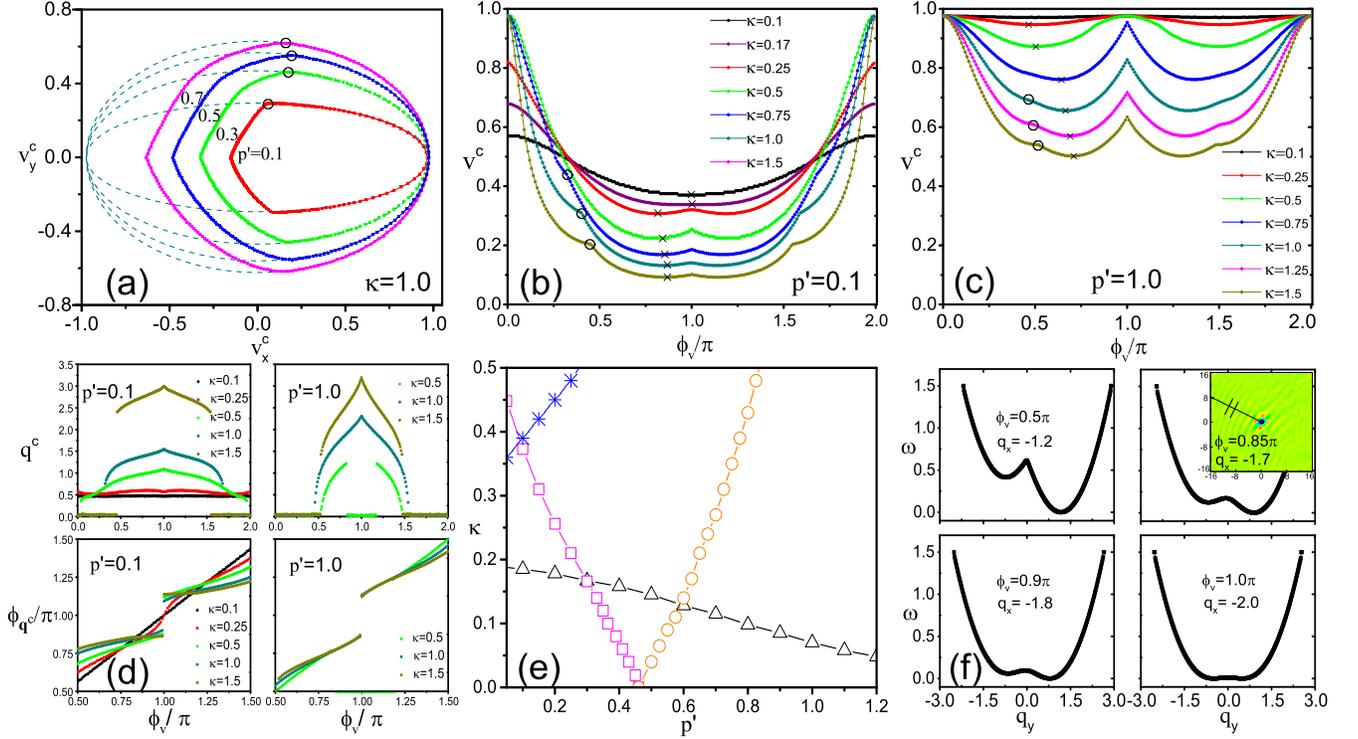}
\caption{(color online) Dependence of the critical excitations(CEs) on the SO coupling and ZF when $\theta_{B}=\pi/2$. (a)The critical velocity $v^{c}$ for different ZF strengths with $\kappa=1.0$. The dashed lines represent the analytical results given by Eq.(11) while the dotted lines represent the numerical ones solving the Bogoliubov equation, where the open circles denote the transition points of the CEs. (b)-(c) $v^{c}$ versus obstacle's moving direction $\phi_{\bm{v}}$ with $p'=0.1$ and $p'=1.0$, respectively. The crosses denote the locations of the minima at $\phi^{m}_{\bm{v}}$. (d) Left(Right) column: Magnitude $q^{c}$ and azimuth angle $\phi_{\bm{q}^{c}}$ of the CE momentum $\bm{q}^{c}$ in (b)((c)). (e)Classification of the ground states. The squares(circles) represent the boundary of the CE when the obstacle is moving along $\bm{x}$(-$\bm{x}$) axis, where the left(right) part of the curve means $\bm{q}^{c}\neq0$($\bm{q}^{c}=0$). For the left part of circles, the stars denote the curve above(below) which $\phi_{\bm{q}^{c}}\neq\pi$($\phi_{\bm{q}^{c}}=\pi$). The triangles give the boundary between the region with $\phi^{m}_{\bm{v}}=\pi$ and that with $\phi^{m}_{\bm{v}}\neq\pi$. Eight regions can thus be identified. (f)Bogoliubov excitation spectrum as a function of $q_{y}$ for different $\phi_{\bm{v}}$. Here $p'=0.1$, $\kappa=1.0$ and $q_{x}$ is fixed to be $q^{c}_{x}$ which is $-1.2, -1.7,-1.8,-2.0$ respectively for different $\phi_{\bm{v}}$. The inset shows the spin density distribution for the corresponding subfigure, where the arrow and two short dashes denote the propagating direction and wavelength of the excited spin wave.  }
\end{figure*}
In Fig.1 we give the excitation spectrum $\omega_{\bm{q}}$ in different situations. The gauge freedom of the ground state indicates that at least one branch of the excitation spectrum always satisfies: $\omega_{\bm{q}=0}=0$(see Fig.1(a1),(b1),(c1)). The stability of the ground states has also been confirmed since all $\omega_{\bm{q}}$ obey $\omega_{\bm{q}}\geq0$. When $p'=\kappa=0$, the three isotropic excitation branches can be given analytically: $\omega^{1}_{\bm{q}}=\sqrt{\frac{q^{2}}{2}(\frac{q^{2}}{2}+2+2\gamma)}$, $\omega^{2}_{\bm{q}}=\frac{q^{2}}{2}$ and $\omega^{3}_{\bm{q}}=\frac{q^{2}}{2}-2\gamma$, where the former one is linear in $\bm{q}$ while the latter two are quadratic in $\bm{q}$ when $\bm{q}\rightarrow0$, as presented in Fig.1(a2). Anisotropy is introduced into the spectrum when the SO coupling is taken into account. One peculiar feature is that the linearity of one branch of the spectrum along $q_{x}$ axis survive even for finite SO coupling $\kappa$ and ZF strength $p'$, as can be seen from Fig.1. In the absence of ZF, or when $\theta_{B}=\pi/2$, this `linear' branch is found to be always connected to the fixed point $(\bm{q},\omega_{\bm{q}})=(\bm{0},0)$, at which it takes a constant slope $\sqrt{1+\gamma}$ along $q_{x}$ axis, while the other two branches are generally shifted for finite $\kappa$ and $p'$. These shifted branches would lead to roton-like excitations as can be seen in Fig.1(a4) and Fig.1(b2). When $\theta_{B}=0$, the lowest-energy branch could exhibit the roton excitation behavior, as can be seen in Fig.1(c2). These roton excitations have already been observed in both scalar and spinor BECs \cite{khamehchi,panjw}.

In the absence of ZF, namely, $p'=0$, when the obstacle is moving along $x$ axis, the CE induced by the moving obstacle can be analyzed by $ v^{c}=\min\{\frac{\omega_{\bm{q}}}{|q_{x}|}\}$ . When $\phi_{\bm{v}}=0$, for a relatively smaller SO coupling $\kappa$, the CE happens at finite $\bm{q}^{c}$, as one of the nonlinear branches dominates(see Fig.1(a3)); while for sufficient large $\kappa$, $\bm{q}^{c}$ equals $0$ and $v^{c}$ is approaching $\sqrt{1+\gamma}$, since the dominant role in the excitation is now replaced by the linear branch(see Fig.1(a4)). On the other hand, when $\phi_{\bm{v}}=\pi$, the CE is always governed by the nonlinear branches, indicating $\bm{q}^{c}$ is always nonzero, which corresponds to roton-like excitation. It can be seen that $v^{c}(\pi)$ is much smaller than $v^{c}(0)$, because the nonlinear branches are shifted leftwards with $\kappa$. As $\kappa$ increases, $v^{c}(\pi)$ also shows nonmonotonic behavior, in comparison with the monotonic $v^{c}(0)$. This is due to the exchange of the two nonlinear branches at about $\kappa=0.3$, which plays the leading role respectively in the CE. These are summarized in Fig.2. The moving trajectory of the CE momentum $\bm{q}^{c}$ with increasing $\kappa$ for $\phi_{\bm{v}}=0$ case would form a closed loop, in contrast with the open curve for the $\phi_{\bm{v}}=\pi$ case, as exhibited in the insets of Fig.2. When the obstacle is moving along directions other than $\phi_{\bm{v}}=0$ or $\pi$, any small velocity of the obstacle can make a finite excitation, and thus the CE is always fixed at $\bm{q}^{c}=0$ and $v^{c}=0$. As an illustration, the spectrum along $q_{y}$ axis is shown in Fig.1(a5), in which the lowest-energy one is quadratic at $q_{y}=0$ but governs the low-energy excitations when the obstacle is moving along $\bm{y}$ direction.

Now we consider the effect of the linear ZF. When it is applied along $\bm{x}$ direction, i.e., $\theta_{B}=\pi/2$, the excitation branch connected to the fixed point $(\bm{0},0)$ keeps linear in $\bm{q}$ when $\bm{q}\rightarrow0$ along any directions, as illustrated in Fig.1(b1)-(b5). This results in the finiteness of $v^{c}$ when $\phi_{\bm{v}}\neq 0$ or $\pi$. The dependence of $v^{c}$ on the moving direction of the obstacle in different situations is shown in Fig.3(a)-(c). For sufficient large SO coupling $\kappa$, $v^{c}$ increases with $p'$, while for fixed ZF strength $p'$, roughly speaking, $v^{c}$ decreases with $\kappa$. When scanning the moving direction $\phi_{\bm{v}}$ from $0$ to $\pi$, there exists a transition point of the CE, where the magnitude $q^{c}$ of $\bm{q}^{c}$ changes abruptly from $0$ to a finite value. These transition points are denoted as the open circles in Fig.3(a). Another novel feature of the CE is that the minimum of $v^{c}$ as a function of $\phi_{\bm{v}}$ occurs at $\phi_{\bm{v}}=\pi$ for smaller SO coupling $\kappa$, while occurs at $\phi_{\bm{v}}\neq\pi$ for relatively larger $\kappa$(see Fig.3(b)-(c)). In Fig.3(d), we show the dependence of $\bm{q}^{c}$ on $\phi_{\bm{v}}$, exhibiting that there is a phonon-dominated ¡®$\bm{q}^{c}=0$¡¯ regime of $\phi_{\bm{v}}$ for sufficient large SO coupling. One point to be noted is that when the obstacle is moving oppositely to $\bm{x}$ direction, i.e., $\phi_{\bm{v}}=\pi$, its critical $\bm{q}^{c}$ can be directed along the same direction with $\bm{v}$, i.e., $\phi_{\bm{q}^{c}}$ equals $\pi$, or along two directions deviated symmetrically from $\pi$. Due to the above features, the ground states in $\theta_{B}=\pi/2$ situation can be classified accordingly, and at least eight regions can be identified, as shown in Fig.3(e). When $\phi_{\bm{v}}$ is near $\pi$, besides the global minimum, the lowest-energy branch of the excitation spectrum develops another local minimum. Both of them are located symmetrically about $q_{x}$ axis and would finally become the global minima when $\phi_{\bm{v}}$ is approaching $\pi$, as exhibited in Fig.3(f).
\begin{figure}[!htb]
\includegraphics[scale=0.535]{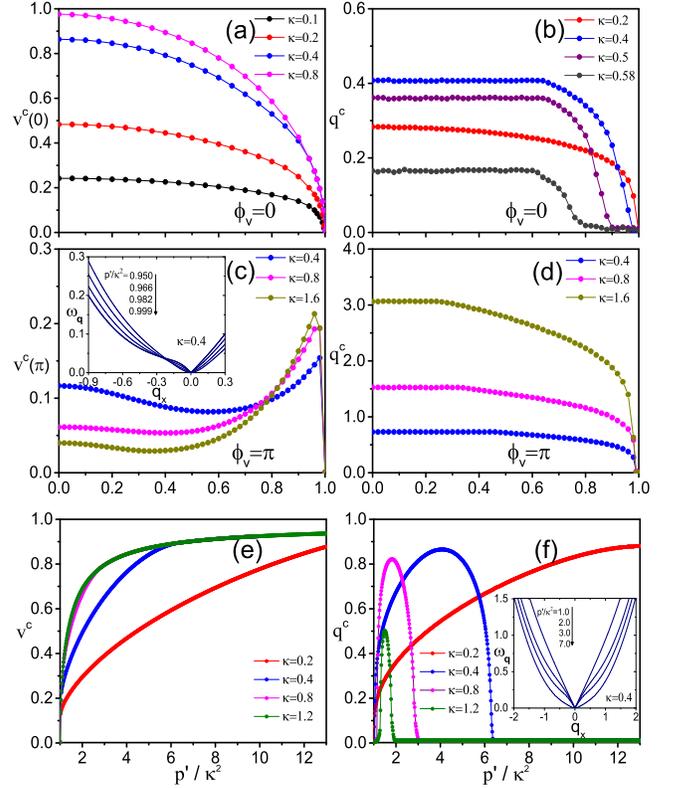}
\caption{(color online) Dependence of the critical velocity $v^{c}(\phi_{\bm{v}})$ and the magnitude of $\bm{q}^{c}$ on the ZF when $\theta_{B}=0$. The left(right) column denotes $v^{c}$($q^{c}$). The upper two rows correspond to $\phi_{\bm{v}}=0$ and $\phi_{\bm{v}}=\pi$ respectively when $p'\leq\kappa^{2}$, while the lower one corresponds to any $\phi_{\bm{v}}$ when $p'\geq\kappa^{2}$, since the system in this situation is isotropic. The insets are the corresponding lowest-energy Bogoliubov excitation spectrum for different $p'/\kappa^{2}$.}
\end{figure}

We now proceed to study the case of $\theta_{B}=0$. As mentioned above, the single-particle ground state is a plane wave with finite momentum $\widetilde{k}=\sqrt{\kappa^{2}-p'^{2}/\kappa^{2}}$ when $p'\leq\kappa^{2}$ while it is static otherwise. The excitation spectrum is shown in Fig.1c(1)-c(5). Except $\phi_{\bm{v}}=0$ and $\pi$ directions, the linearity of the linear branch is destroyed, similar to that of $p'=0$ with SO coupling. Thus the critical velocity $v^{c}(\phi_{\bm{v}})$ is found to be finite only for $\phi_{\bm{v}}=0$ or $\pi$. The critical $\bm{q}^{c}$ is also found to be parallel to the obstacle velocity $\bm{v}$ for the two moving directions. When $p'/\kappa^{2}<1$, for fixed SO coupling, $v^{c}(0)$ decreases monotonically with $p'/\kappa^{2}$, while $v^{c}(\pi)$ shows non-monotonic behavior, as exhibited in Fig.(4)(a)-(d). $v^{c}(\pi)$ forms a peak near $p'/\kappa^{2}=1$ and then decreases sharply to zero when $p'/\kappa^{2}$ is approaching $1$. This is because that despite the left part of the lowest-energy Bogoliubov excitation curve is monotonic, it generally has an inflection point at a finite $q_{x}$, which disappears when $p'/\kappa^{2}$ is approaching $1$, as exhibited in the inset of Fig.4(c). When $p'/\kappa^{2}>1$, the ground state and thus the excitation spectrum become isotropic, and $v^{c}$ monotonically increases with $p'/\kappa^{2}$. The critical $\bm{q}^{c}$ shares the same direction with $\bm{v}$, and for sufficient large $\kappa$, it is always fixed at $\bm{0}$. These are shown in Fig.4(e)-(f). The behavior of $\bm{q}^{c}$ can be understood in a similar way as discussed above from the Bogoliubov excitation curves, as exhibited in the inset of Fig.4(f).
\subsection*{B:Vortex excitations and vortex structures}

\begin{figure*}[!htb]
\includegraphics[scale=0.22]{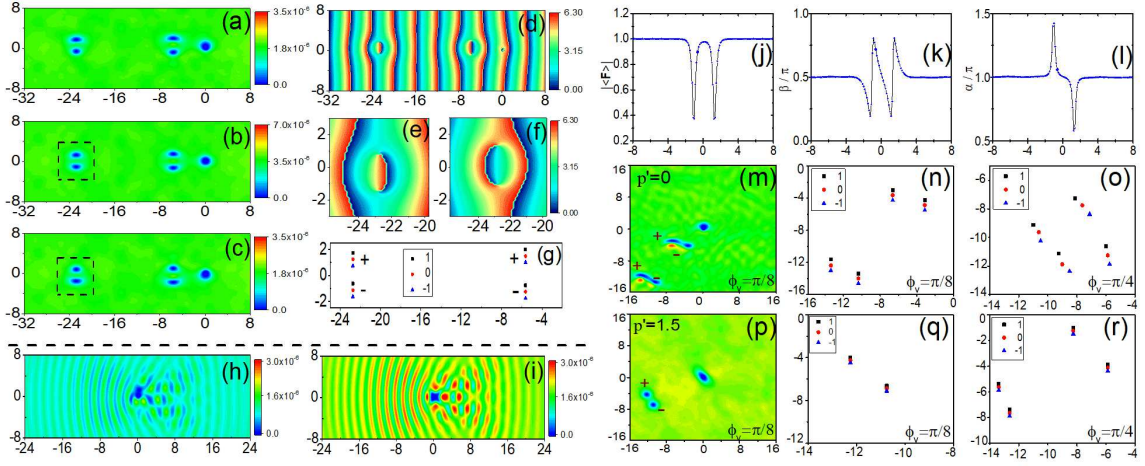}
\caption{(color online) Left two columns: Density plots for (a) $|\psi_{1}|^{2}$, (b) $|\psi_{0}|^{2}$, (c) $|\psi_{-1}|^{2}$ together with the corresponding phase profiles in (d)-(f) when $\theta_{B}=\pi/2$. Note that (e)(f) correspond to the magnifications of $m_{F}=0$ and $m_{F}=-1$ components denoted by dashed squares in (b)(c)respectively. (g) Vortex core positions of the three components at a certain time. Here $(v,\phi_{\bm{v}}) =(0.55,0)$. (h)(i) Density distributions of $m_{F}=1$ and $m_{F}=0$ components respectively at $(v,\phi_{\bm{v}}) =(0.05,\pi)$. Here $\kappa=1.6$, and $p'=0.75$($p'=0$) for the panels above(below) the horizontal dashed line. Right upper row: Magnitude of the normalized spin density $|\langle\bm{F}(\bm{r})\rangle|$, and its Euler angles $\beta$ and $\alpha$ along the vortex chain in the dashed square region in (b). Right two lower rows: Density distribution of $m_{F}=0$ component when $(v,\phi_{\bm{v}})=(0.3,\frac{\pi}{8})$ for (m) and $(0.6,\frac{\pi}{8})$ for (p). Positions of the three components of vortex dipoles with $(v,\phi_{\bm{v}})$ being $(0.3,\frac{\pi}{8})$ for (n), $(0.3,\frac{\pi}{4})$ for (o), $(0.6,\frac{\pi}{8})$ for (q), $(0.6,\frac{\pi}{4})$ for (r) respectively. Here $\kappa=1.6$, and $p'=0$($p'=1.5$) is for the right middle(bottom) row.}
\end{figure*}

In this section, we give our numerical results of our time-dependent GP equation. Compared with the ideal case discussed above, for a finite-size obstacle, the Bogoliubov excitations can be expected to be qualitatively unchanged, but the critical velocity $v^{c}$ generally decreases\cite{stie,kwon2,kato3}. Besides Bogoliubov excitations in momentum space, when the spinor BEC is disturbed by a moving obstacle, vortex excitations in real space can be induced. Since the vortex street phenomenon is very sensitive to the parameter chosen \cite{sasaki,kwona}, here we are only focused on the process of vortex-antivortex pair generation. In the following discussion, for simplicity, the width of the obstacle is fixed to be $R=0.4$. Simulations of Eq.(8) are performed by using a Fourier pseudospectral split-step method, as well as the fourth-order Runge-Kutta scheme. The initial state we choose is the ground state as the obstacle is static, which becomes a plane-wave state far away from the obstacle. This state is prepared by the imaginary-time evolution method, in which $i$ on the left-hand side of Eq.(8) is replaced with $-1$. The numerical simulations presented here are performed in a grid of $512\times512$ points, with a lattice spacing of $1/8$ in both directions.

First, we consider the case of $\theta_{B}=\pi/2$, i.e., the ZF is applied along $\bm{x}$ direction. Vortex excitations could be shed from the fast moving obstacle\cite{kwonb, kato3}. When the obstacle is also moving along $\bm{x}$ direction, vortex-antivortex pairs are generated periodically behind it for each component. The vortex and antivortex cores for each pair are aligned vertically for a sufficient large SO coupling, as shown in Fig.5(a)-(f). The three vortex(antivortex) cores for the three components are also displaced and arranged alternately in a vertical line, which can be regarded as a whole as a coreless vortex(antivortex), as can be seen in Fig.5(g). When the obstacle is moving oppositely to $\bm{x}$ direction, an unusual phenomenon occurs when the SO coupling become weak, which gives rise to a rather small critical velocity $v^{c}$ (see Fig.1(b2)-(b4)). Here, Bogoliubov spin excitations are excited by the slowly moving obstacle. The spin waves in $m_{F}=1$ and $m_{F}=-1$ components share the exactly similar patterns(forming density peaks and valleys at the same locations) with their densities aligning alternatively with $m_{F}=0$ component, as shown in Fig.5(h)-(i). Slightly increase of the obstacle velocity along $-\bm{x}$ direction will cause turbulence. The generation of vortex-antivortex pairs is also possible, but is sensitive to parameters.

To study the vortex structure in detail, we analyze the normalized spin density defined by $\langle\bm{F}(\bm{r})\rangle={\xi}^{*}(\bm{r})\bm{\widehat{F}}\xi(\bm{r})$ with $\xi$ given by $\xi=\Psi/\sqrt{n}$. For a coreless vortex in a FM spinor BEC without SO coupling, its magnitude is unity everywhere, i.e., $|\langle\bm{F}(\bm{r})\rangle|=1$ \cite{love2012}. Actually, a general FM spinor can be constructed by a spin rotation $U(\alpha,\beta,\gamma)=e^{-iF_{z}\alpha}e^{-iF_{y}\beta}e^{-iF_{z}\gamma}$
defined by the three Euler angles $\alpha$, $\beta$, $\gamma$, acting on a spinor pointing to $\bm{z}$,
\begin{equation}
\xi= e^{i\delta}U(\alpha,\beta,\gamma)
\left (
\begin{array}{c}
1\\
0\\
0
\end{array}
\right )
=e^{-i\gamma^{'}}
\left (
\begin{array}{c}
e^{-i\alpha}\cos^{2}\frac{\beta}{2}\\
\frac{1}{\sqrt{2}}\sin\beta\\
e^{i\alpha}\sin^{2}\frac{\beta}{2}
\end{array}
\right ),
\end{equation}
where $\gamma^{'}=\gamma-\delta$ with $\delta$ the global phase of the condensate. However, the normalized spin density of the coreless vortex we study here cannot be described by an FM spinor. In our situation, $|\langle\bm{F}(\bm{r})\rangle|\neq1$, especially near the vortex cores. Actually, the Euler angles $\alpha$ and $\beta$ can still be introduced and be defined straightforwardly from $\langle\bm{F}(\bm{r})\rangle\equiv|\langle\bm{F}(\bm{r})\rangle|(\sin\beta\cos\alpha,\sin\beta\sin\alpha,\cos\beta)$. Away from the vortices and obstacle, the system is well described by the plane-wave FM ground state, indicating that $\alpha$ and $\beta$ are approaching $\pi$ and $\pi/2$ respectively. When scanning along a vertical line across the vortex-antivortex pairs, $|\langle\bm{F}(\bm{r})\rangle|$ forms valleys at exactly the locations of the vortex and antivortex cores of $m_{F}=0$ component, with $\alpha$ and $\beta$ modulating around $\pi$ and $\pi/2$ respectively, as exhibited in Fig.5(j)-(l). When away from the vortex(antivortex) core, the coreless vortex can be well described by
\begin{equation}
 \xi(\bm{r})\propto e^{\pm i\theta(\bm{r})}
\left (
\begin{array}{c}
\frac{1}{2}\\
-\frac{\sqrt{2}}{2}\\
\frac{1}{2}
\end{array}
\right ),
\end{equation}
with $\theta(\bm{r})$ being the azimuth angle of $\bm{r}$. Compared with the vortex with topological skyrmion structure in a SO coupled FM BEC generated by rotation\cite{xxq1,lcf1}, this kind of vortex or antivortex generated by a moving obstacle is topologically trivial. Simulations for the obstacle moving in other directions are also performed. The induced vortex-antivortex pairs are no longer arranged vertically, but the three vortex (antivortex) cores for the three components could still keep vertical under strong enough ZF. These are shown in Fig.5(m)-(r).

\begin{figure}[!htb]
\includegraphics[scale=0.61]{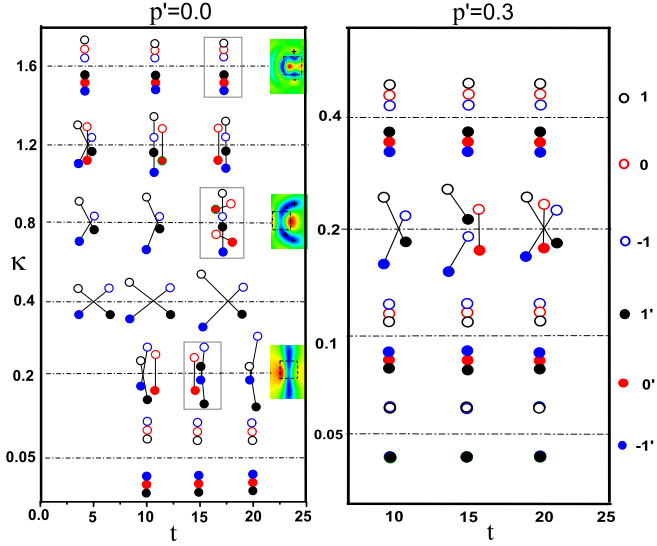}
\caption{(color online) Time evolution of the vortex and antivortex structure when $\theta_{B}=\pi/2$ for different SO coupling strengths. The hollow (solid) circles denote the relative locations of vortices (anti-vortices) for the three components. The left (right) column corresponds to $p'=0$ ($p'=0.3$). The insets denote the corresponding density distributions of $m_{F}=0$ component in the regions marked by the gray squares.}
\end{figure}

Now we investigate the dependence of the vortex structure on the SO coupling and ZF, as well as its stability under time evolution. We found that the vertical vortex(antivortex) structure can keep stable under time evolution for relatively strong or weak SO coupling. For each coreless vortex(antivortex), the vortex cores for the three components are arranged in a vertical line, and both the vortex and antivortex share the exactly same order of the cores, i.e., in $1$, $0$, $-1$ order from top to bottom for a relatively strong SO coupling. Similar linear structure has also been found in the coreless vortex in the rotating SO coupled spin-1 BEC\cite{chain}. Nevertheless, interestingly, both the orders would be reversed for the weak SO coupling. For the intermediate SO coupling, the spatial arrangement of the cores for the three components is non-collinear and also vary with time. The width of this non-collinear region is suppressed as the strength of the ZF increases. These are exhibited in Fig.6.

\begin{figure}[!htb]
\includegraphics[scale=0.23]{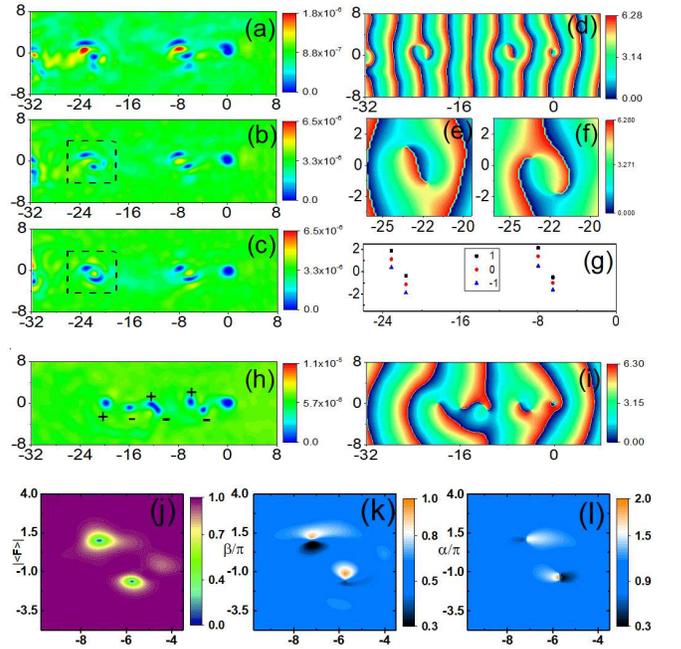}
\caption{(color online) (a)-(g): The same as the left panels in Fig.5 except $\theta_{B}=0$, $p'=0.48\kappa^{2}$, $\kappa=1.6$ and $(v,\phi_{\bm{v}}) = (0.4,0)$. (h)-(i) Density and phase profiles for $m_{F}=-1$ component at $(v,\phi_{\bm{v}}) = (0.4,0)$ with $p'=0.93\kappa^{2}$. (j)-(l): Spatial distribution of the magnitude of the normalized spin density $|\langle\bm{F}(\bm{r})\rangle|$, together with its Euler angles $\beta$ and $\alpha$ near the vortex dipoles.}
\end{figure}

Secondly, we turn to study the vortex shedding for the case of $\theta_{B}=0$, i.e., the ZF is applied along $\bm{z}$ direction. When $p'<\kappa^{2}$, the vortex-antivortex pairs generated by the moving obstacle along $\bm{x}$ direction become canted from the vertical line, as can be seen in Fig.7(a)-(f). The canted angle increases with the ZF strength(Fig.7(h)-(i)). However, for each vortex or antivortex, the cores for the three components still keep vertical, as schematically shown in Fig.7(g) by the locations of the cores. The normalized spin density around a vortex(antivortex) shows similar pattern. $|\langle\bm{F}(\bm{r})\rangle|$ forms valleys and the spin directions vary dramatically around the vortex core, as can be seen in Fig.7(j)-(l). The structure of this kind of vortex(antivortex) is still topologically trivial.

\begin{figure}[!htb]
\includegraphics[scale=0.20]{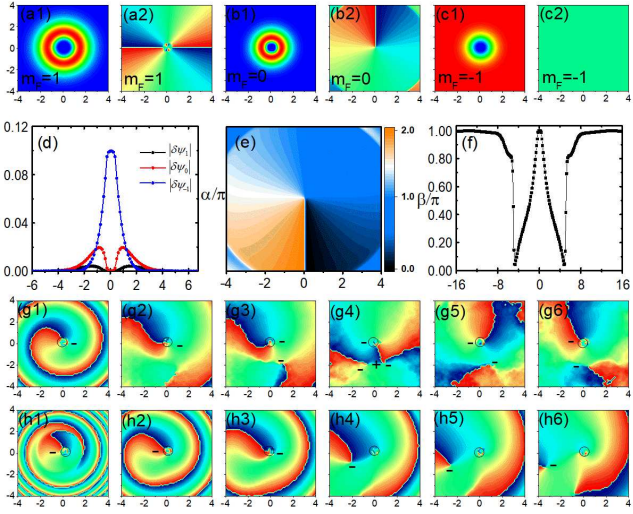}
\caption{(color online) Skyrmion-like vortex induced and trapped by the obstacle when $\theta_{B}=0$ and $p'>\kappa^{2}$. (a)-(c): Density and phase profiles for the three components in the ground state. (d)-(f): Magnitude of deviations $\delta\psi_{1}$, $\delta\psi_{0}$ and $\delta\psi_{-1}$  induced by the obstacle, as well as the Euler angles $\alpha$ and $\beta$. Here $\kappa=0.8$ and $p'=1.01\kappa^{2}$. Time evolution of the phase profiles for $m_{F}=0$ component at $t=$0.4, 2, 4, 5, 10, 20 in (g1)-(g6) with $v=0.2$ respectively, while at $t=$0.2, 0.4, 2, 4, 6, 8 in (h1)-(h6) with $v=0.6$ respectively. The open circles denote the position of the obstacle and the symbols $+(-)$ denote the induced vortices(antivortices). }
\end{figure}
When $p'>\kappa^{2}$, the ground state in the absence of the obstacle is the fully spin polarized state $(0,0,1)^{T}$, where all the atoms have been condensed into $m_{F}=-1$ component. Due to the finite SO coupling, the introduction of the obstacle will inevitably induce atoms around the obstacle in $m_{F}=0$, $1$ components. Remarkably, we find a skyrmion-like ground state in the presence of the obstacle, as shown in Fig.8(a1)-(f). Effectively, the static obstacle has actually induced and trapped an antivortex in $m_{F}=0$ component and a multiply quantized antivortex with winding number $-2$ in $m_{F}=1$ component. When away from the center of the obstacle, namely, when $r$ is larger than several $\xi$, this skyrmion-like ground state can be approximately described by
\begin{equation}
\xi(\bm{r})\propto
\left (
\begin{array}{c}
-e^{-i2\theta(\bm{r})}\cos^{2}\frac{\beta(r)}{2}\\
-i\frac{1}{\sqrt{2}}e^{-i\theta(\bm{r})}\sin\beta(r)\\
\sin^{2}\frac{\beta(r)}{2}
\end{array}
\right ), \beta(r)\rightarrow\pi.
\end{equation}
Here the Euler angle $\alpha$ can be expressed as $\alpha(\bm{r})=\theta(\bm{r})+\pi/2$, while $\beta(r)$ is changing from $\pi$ to $0$ when $r$ is approaching to the center of the skyrmion, but this process is truncated by the presence of the obstacle. This topological ground state is even found to be dynamically stable. When the obstacle is moving at a relatively small velocity, this topological structure is still maintained(see Fig.8(g1)-(g6)). A sufficiently large obstacle velocity would lead to the destruction of this structure, and the trapped antivortex or multiply quantized antivortex in $m_{F}=0$ or $m_{F}=1$ component would finally escape from the obstacle under time evolution, as exhibited in Fig.8(h1)-(h6).

\section{SUMMARY}

In summary, we have studied the dynamics of a SO coupled spin-1 BEC flow past an obstacle in the presence of a linear ZF. First, we have analyzed the Bogoliubov excitation spectrum and demonstrated the dependence of the critical velocity and corresponding critical excitation momentum on the obstacle's moving direction, as well as the strengthes of the SO coupling and ZF. When the ZF is applied along $\bm{x}$ direction, we find the ground states can be classified into eight regions in parameter space. When the ZF is exerted along $\bm{z}$ direction, the ground state and its excitation spectrum is found to exhibit anisotropic behavior if the ZF is weak while become isotropic if the ZF is sufficiently strong. Secondly, by solving the time-dependent GP equation, we have numerically investigated the combined effect of the SO coupling and ZF on the vortex structure. In the case of the ZF being along $\bm{x}$ direction, under a weak or strong SO coupling, the vortex cores of the three components are displaced and arranged in a vertical line, where their order would be reversed by changing the SO coupling. In the case of the ZF being along $\bm{z}$ direction, we find that a skyrmion-like vortex ground state could be induced even by a static obstacle, and if the obstacle is moving at a relatively small velocity, this topological structure exhibits dynamical stability.

\section*{ACKNOWLEDGMENTS}
Q. L. Z. thanks W. P. Chen, Y. Zhou, F. Xiong and J. P. Xiao for useful discussions.
This work is supported by NSFC Project No. 111774126 and 973 Project No. 2015CB921202.

\section{appendix}
In this appendix we show in detail the Bogoliubov equation when the ZF is taken along $\bm{x}$ ($\theta_{B}=\pi/2$) or $\bm{z}$ ($\theta_{B}=0$) direction respectively. Assuming the ground state is $e^{ i\tilde{\kappa} x}(\alpha_{1},\alpha_{0},\alpha_{-1})^{T}$, where $\tilde{\kappa}=\kappa(\tilde{\kappa}=\sqrt{\kappa^{2}-p^{'2}/\kappa^{2}})$ for $\theta_{B}=\pi/2$($\theta_{B}=0$). According to Bogoliubov theory, $i\hbar\partial_{t}\delta\Psi(\bm{r})= (\mathcal{H}+i\bm{v}\cdot\nabla)\delta\Psi(\bm{r})+\mathcal{H}^{'}\delta\Psi^{*}(\bm{r})$, where $\mathcal{H}=\mathcal{H}_{kin}+\mathcal{H}_{int}$ with

\begin{widetext}
\begin{equation}
\mathcal{H}_{kin}=
\left (
\begin{split}
\begin{array}{ccc}
-\frac{1}{2}(\nabla+\bm{i}\tilde{\kappa})^{2}+p'_{z}-\mu          & \frac{\kappa}{\sqrt{2}}(\tilde{\kappa}-i\partial_{-})+ \frac{p^{'}_{x}}{\sqrt{2}}     & 0  \\
\frac{\kappa}{\sqrt{2}}(\tilde{\kappa}-i\partial_{+})+ \frac{p^{'}_{x}}{\sqrt{2}}  &-\frac{1}{2}(\nabla+\bm{i}\tilde{\kappa})^{2}-\mu  &\frac{\kappa}{\sqrt{2}}(\tilde{\kappa}-i\partial_{-})+ \frac{p^{'}_{x}}{\sqrt{2}} \\
0                                    & \frac{\kappa}{\sqrt{2}}(\tilde{\kappa}-i\partial_{+})+ \frac{p^{'}_{x}}{\sqrt{2}}      &-\frac{1}{2}(\nabla+\bm{i}\tilde{\kappa})^{2}-p'_{z}-\mu
\end{array}
\end{split}
\right ) ,
\end{equation}

\begin{equation}
\begin{small}
\mathcal{H}_{int}=
\left (
\begin{array}{ccc}
1+|\alpha_{1}|^{2}+\gamma(2|\alpha_{1}|^{2}+|\alpha_{0}|^{2}-|\alpha_{-1}|^{2}) & \alpha_{0}^{*}\alpha_{1}+\gamma(\alpha_{0}^{*}\alpha_{1}+2\alpha_{-1}^{*}\alpha_{0}) &
(1-\gamma)\alpha_{-1}^{*}\alpha_{1} \\
\alpha_{1}^{*}\alpha_{0}+\gamma(\alpha_{1}^{*}\alpha_{0}+2\alpha_{0}^{*}\alpha_{-1})  &1+|\alpha_{0}|^{2}+\gamma(|\alpha_{1}|^{2}+|\alpha_{-1}|^{2})                &\alpha_{-1}^{*}\alpha_{0}+\gamma(\alpha_{-1}^{*}\alpha_{0}+2\alpha_{0}^{*}\alpha_{1}) \\
(1-\gamma)\alpha_{1}^{*}\alpha_{-1}              &\alpha_{0}^{*}\alpha_{-1}+\gamma(\alpha_{0}^{*}\alpha_{-1}+2\alpha_{1}^{*}\alpha_{0})          &1+|\alpha_{-1}|^{2}+\gamma(2|\alpha_{-1}|^{2}+|\alpha_{0}|^{2}-|\alpha_{1}|^{2})
\end{array}
\right ),
\end{small}
\end{equation}

and

\begin{equation}
\mathcal{H}^{'}=
(1+\gamma) \left(
\begin{array}{c}
\alpha_{1}\\
\alpha_{0}\\
\alpha_{-1}
\end{array}
\right)(\alpha_{1}, \alpha_{0},\alpha_{-1}).
\end{equation}

In momentum $\bm{q}$-space, $\mathcal{H}$ can be expressed as,
\begin{equation}
\mathcal{H}_{\bm{q}}=
\left (
\begin{split}
\begin{array}{ccc}
\frac{1}{2}(q^{2}+2q_{x}\tilde{\kappa}+\tilde{\kappa}^{2})+p'_{z}-\mu          & \frac{\kappa}{\sqrt{2}}(\tilde{\kappa}+q_{-})+ \frac{p^{'}_{x}}{\sqrt{2}}     & 0  \\
\frac{\kappa}{\sqrt{2}}(\tilde{\kappa}+q_{+})+ \frac{p^{'}_{x}}{\sqrt{2}}  &\frac{1}{2}(q^{2}+2q_{x}\tilde{\kappa}+\tilde{\kappa}^{2})-\mu  &\frac{\kappa}{\sqrt{2}}(\tilde{\kappa}+q_{-})+ \frac{p^{'}_{x}}{\sqrt{2}} \\
0                                    & \frac{\kappa}{\sqrt{2}}(\tilde{\kappa}+q_{+})+ \frac{p^{'}_{x}}{\sqrt{2}}      &\frac{1}{2}(q^{2}+2q_{x}\tilde{\kappa}+\tilde{\kappa}^{2})-p'_{z}-\mu
\end{array}
\end{split}
\right )+\mathcal{H}_{int},
\end{equation}
\end{widetext}
where $q_{\pm}=q_{x} \pm iq_{y}$.

When $\theta_{B}=\pi/2$, $\alpha_{1}=\alpha_{-1}=1/2$, $\alpha_{0}=-\sqrt{2}/2$, and the chemical potential $\mu=-\frac{\kappa^{2}}{2}+1+\gamma$, which means that :

\begin{equation}
\mathcal{H}_{int}=\frac{1}{4}
\left(
\begin{array}{ccc}
5+3\gamma              & -\sqrt{2}(1+3\gamma)     &  1-\gamma \\
-\sqrt{2}(1+3\gamma)    &   2(3+\gamma)            &  -\sqrt{2}(1+3\gamma) \\
1-\gamma               & -\sqrt{2}(1+3\gamma)     &  5+3\gamma
\end{array}
\right ),             
\end{equation}

and
\begin{equation}
\mathcal{H}^{'}=\frac{(1+\gamma)}{4}
\left (
\begin{array}{ccc}
1         &-\sqrt{2}   & 1         \\
-\sqrt{2} &2           &-\sqrt{2}        \\
1         &-\sqrt{2}   & 1
\end{array}
\right ), \\  
\end{equation}

For $\theta_{B}=0$, when $p'<\kappa^{2}$, $\alpha_{1}=\frac{1}{2}-\frac{p'^{2}}{2\kappa^{4}}$, $\alpha_{-1}=\frac{1}{2}+\frac{p'^{2}}{2\kappa^{4}}$ and $\alpha_{0}=-\sqrt{2\alpha_{1}\alpha_{-1}}$ with the chemical potential $\mu=-\frac{\kappa^{2}}{2}+1+\gamma-\frac{p'^{2}}{2\kappa^{2}}$. The expressions for $\mathcal{H}_{int}$ and $\mathcal{H}^{'}$ given by Eq.(17)-(18) can hardly be simplified. When $p'>\kappa^{2}$,
$\alpha_{1}=\alpha_{0}=0$, $\alpha_{-1}=1$,
\begin{equation}
\mathcal{H}_{int}=
\left(
\begin{array}{ccc}
1-\gamma        &       0               &  0                    \\
0               &   1+\gamma            &  0                    \\
0               &       0               &  2(1+\gamma)
\end{array}
\right ),             
\end{equation}
and
\begin{equation}
\mathcal{H}^{'}=
\left (
\begin{array}{ccc}
0         &0        & 0        \\
0         &0        & 0        \\
0         &0        & 1+\gamma
\end{array}
\right ), \\  
\end{equation}
with the chemical potential $\mu = 1+\gamma- p'$.

\end{document}